\documentclass[floatfix,reprint, superscriptaddress, amsmath,amssymb, aps]{revtex4-2}

\usepackage{graphicx} 
\usepackage[utf8]{inputenc}
\usepackage[T1]{fontenc}
\usepackage{hyperref}
\usepackage{lmodern}
\hypersetup{colorlinks=true,allcolors=blue}

\usepackage{braket}
\newcommand{\gtwo}{g^{(2)}}
\newcommand{\tudo}{Condensed Matter Theory, TU Dortmund, Otto-Hahn-Str. 4, 44227 Dortmund, Germany}

\date{October 2023}

\begin{document}

\title{Few-Photon SUPER: Quantum emitter inversion via two off-resonant photon modes}

\author{Quentin W. Richter}
\affiliation{%
\tudo
}%
\author{Jan M. Kaspari}
\affiliation{%
\tudo
}%
\author{Thomas K. Bracht}%
\affiliation{%
\tudo
}%
\author{Leonid Yatsenko}
\affiliation{Institute of Physics, National Academy of Science of Ukraine, Nauky
Avenue 46, Kyiv 03028, Ukraine}

 \author{Vollrath Martin Axt}
\affiliation{Lehrstuhl für Theoretische Physik III, Universität Bayreuth, 95440 Bayreuth, Germany}
 \author{Arno Rauschenbeutel}
\affiliation{Department of Physics, Humboldt-Universität zu Berlin, 10099 Berlin, Germany}
 \author{Doris E. Reiter}
\affiliation{%
\tudo
}%

\begin{abstract}
With the realization of controlled quantum systems, exploring excitations beyond the resonant case opens new possibilities. We investigate an extended Jaynes-Cummings model where two non-degenerate photon modes are coupled off-resonantly to a quantum emitter. This allows us to identify few-photon scattering mechanisms that lead to a full inversion of the emitter while transferring off-resonant photons from one mode to another. This behavior connects to recent measurements of a two-level emitter scattering two off-resonant photons simultaneously. Furthermore, our results can be understood as quantized analogue of the recently developed off-resonant quantum control scheme known as Swing-UP of quantum EmitteR (SUPER). Our intuitive formalism gives a deeper insight into the interaction of a two-level emitter with off-resonant light modes with the prospect of novel photonic applications.
\end{abstract}

\maketitle

\section{Introduction}

The two-level emitter and its interaction with light lies at the heart of quantum optics. The most prominent case is when the frequency of the light field is in resonance with the transition frequency of the two-level system, leading to phenomena such as vacuum Rabi oscillations \cite{rabi1937space,reithmaier2004strong,khitrova2006vacuum} and the Mollow triplet \cite{mollow1969power}. Adding a chirp to the exciting laser pulse leads to adiabatic rapid passage \cite{landau1932theory,zener1932non}. 
From textbooks we learn that for a two-level emitter that is excited with a single-frequency field, a full inversion can only be achieved in resonance, while a detuned single-frequency pulse cannot lead to a full inversion. 

This changes drastically, when instead of a single off-resonant pulse, multi-color pulses that are all off-resonant are used \cite{guerin2001dynamical, conde2005experimental, bracht2021swing,koong2021coherent,he2019coherently, karli2022super, boos2024coherent, joos2023triggered, torun2023super}. Alternatively, a frequency modulation of an off-resonant pulse has similar effects \cite{vitanov2003population,gagnon2017suppression,bracht2021swing,shi2021}. An intuitive explanation of the inversion process can be given by the Swing-UP of quantum EmitteR (SUPER) mechanism \cite{bracht2021swing}. The SUPER mechanism has also been demonstrated experimentally  \cite{karli2022super, boos2024coherent, joos2023triggered, torun2023super}. In the most widely discussed case, a pair of two red-detuned pulses, i.e., both pulses have an energy below the transition frequency, are employed to excite a two-level system. This prompts the question of energy conversion: In a single photon picture, neither of the pulses would have enough energy to excite the two-level system, while also their sum or difference does not match the resonance condition. 

In this paper, we clarify this question by considering an intuitive model of an extended Jaynes-Cummings model with two photon modes. We find distinct resonance conditions which lead to an inversion of the two-level emitter coupled to two off-resonant, non-degenerate modes. At these resonances, a multi-photon scattering process takes place.

Understanding off-resonant excitation schemes is not only essential for the generation of non-classical single photon states for quantum communication \cite{senellart2017high,vajner2022quantum}, but also interesting for understanding the fundamentals of the interaction of a two-level emitter with photons \cite{le2022dynamical,masters2023simultaneous}. The interaction of a single two-level emitter with few photons, either resonant or off-resonant, results in a multi-photon scattering, visible in the correlation function \cite{dalibard1983correlation,aspect1980time}. The $\gtwo$-correlation function is a standard measure for photon-photon correlations that occur in resonance fluorescence \cite{kimble1977resonance}. In Ref.~\cite{aspect1980time, masters2023simultaneous} it was shown, that a correlated photon scattering takes place. In the far off-resonant case, this will lead to the excitation of the two-level emitter in a three-photon process, where two drive photons disappear, while one photon appears at even larger detuning. Here, we connect these multi-photon scatterings to the SUPER scheme.

Following this introduction, we will give a brief repetition of the excitation of a two-level emitter with either a single monochromatic pulse resulting in Rabi oscillations or using two-color excitation via the SUPER scheme in the semiclassical picture in Sec.~\ref{sec:semiclassical}. Next we introduce the two-mode Jaynes-Cummings model in Sec.~\ref{sec:two-mode-JC}. Within the two-mode Jaynes-Cummings model we discuss the dynamics of the participating states and the maximally achievable occupation of the two-level emitter to identify a quantum analog of SUPER in Sec.~\ref{sec:dynamics}, before discussing and summarizing our findings in Sec.~\ref{sec:conclusion}.
%
%
\section{Rabi and SUPER in the semiclassical picture}
\label{sec:semiclassical}

\begin{figure}[htb]
    \centering
    \includegraphics[width=\columnwidth]{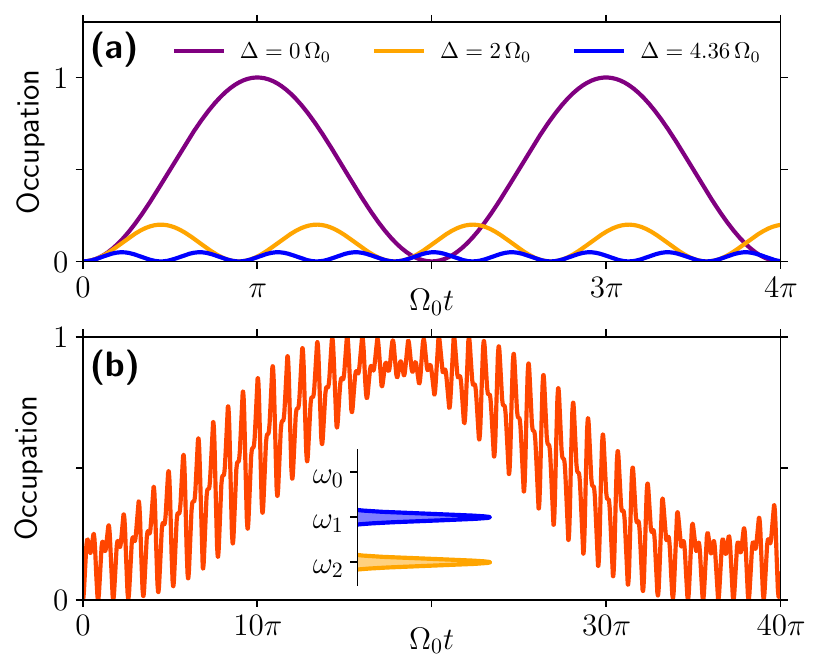}
    \caption{\textbf{Rabi and SUPER scheme in the semiclassical picture}: (a) Rabi oscillations in the occupation of the excited states as a function of time for the resonant case (purple), and the off resonant cases with detunings $\Delta_1$ (blue), $\Delta_2$ (yellow). (b) Swing-Up dynamics of the occupation induced by using two pulses with constant amplitude $\Omega_0$ and different detunings $\Delta_1 = 2\,\Omega_0$ and $\Delta_2 = 4.36\,\Omega_0$. The pulses each have a duration of $15.16\pi/\Omega_0$. The inset indicates the central frequencies of the two exciting lasers. }
    \label{fig:semi}
\end{figure}

We consider a two-level emitter consisting of the ground state $|g\rangle$ and the excited state $|x\rangle$ with  energy difference $E_x-E_g=\hbar\omega_0$ coupled to the electric field in dipole approximation, with resonant or close-to-resonant excitation, such that the rotating wave approximation is applicable. 

In the context of optical excitation schemes of quantum emitters, usually a semi-classical picture via the standard Hamiltonian is employed, reading
\begin{equation}
    H = \hbar\omega_0\ket{x}\bra{x}-\frac{\hbar}{2}\left(\Omega_{\text{laser}}(t)\ket{x}\bra{g}+h.c.\right). \label{eq:hamiltonian_2ls}
\end{equation}
In the single field case, the coupling to the light field is given by
\begin{equation}
    \Omega^{\text{single}}_{\text{laser}}(t) = \Omega(t)e^{-i\omega_L t } .
    \label{eq:1puls} 
\end{equation}
The electric field $E$ is contained in the Rabi frequency $\Omega=2dE/\hbar$ via its coupling to the dipole moment $d$. Within the rotating wave approximation we use a notation splitting the field into the complex-valued part oscillating with the frequency $\omega_L$ and the real-valued envelope $E(t)\sim\Omega(t)$. The detuning is then defined as the difference of the laser frequency and the transition frequency of the two-level system, $\Delta = \omega_L-\omega_0$. An cw-excitation switched on at $t=0$ leads to Rabi oscillations in the occupation of the excited state as displayed in Fig.~\ref{fig:semi}(a).

A two-color light field contains two frequencies $\omega_i$, such that the coupling to the laser reads
\begin{equation}
    \Omega^{\text{two-color}}_{\text{laser}}(t) = \Omega_1(t)e^{-i\omega_1 t } + \Omega_2(t)e^{-i\omega_2 t} \,. 
    \label{eq:pulses} 
\end{equation} 
Besides the resonant case, the two-level system can be excited with an off-resonant pulse pair, i.e.,  $\omega_1,\omega_2\neq\omega_0$ or in terms of detunings $\Delta_1,\Delta_2\neq 0$. In  Ref.~\cite{bracht2021swing} we numerically found that an excitation with a Gaussian pulse pair leads to a complete inversion if the detunings fulfil the resonance condition 
\begin{equation} \label{eq:resonance_2CSUPER}
    \Delta_2 = \Delta_1 + \sqrt{\Delta_1^2+ \left(\Omega_1^{\text{max}}\right)^2},
\end{equation}
with $\Omega_1^{\text{max}}$ being the maximal pulse strength.  
Employing a semiclassical dressed-state picture gives further insight into the resonance condition \cite{bracht2023dressed}.

For the case of two off-resonant fields with the same constant amplitude $\Omega_0$ switched on at $t=0$, shown in Fig.~\ref{fig:semi}(b), we find complete inversion for the condition
\begin{equation} \label{eq:resonance_CWSUPER}
    \sqrt{\Omega_0^2 +\Delta_2^2} = 2\sqrt{\Omega_0^2 + \Delta_1^2} \quad \text{or} \quad
    \Omega_{R,2} = 2\Omega_{R,1}
\end{equation}
with $\Omega_{R,i}$ being the respective Rabi frequency. The corresponding occupation dynamics shows a behavior that is typical for SUPER \cite{bracht2021swing}. Note that here both detunings are negative, i.e., $\omega_{1,2} < \omega_0$. The same dynamics occur for both detunings being positive $ \omega_{1,2} > \omega_0$. We remark that the two resonant conditions Eq.~\eqref{eq:resonance_2CSUPER} and Eq.~\eqref{eq:resonance_CWSUPER} are for two different excitation conditions, namely pulsed excitation, where the instantaneous Rabi frequency changes along the pulse, and two constant fields switched on instantaneously with the same amplitude, respectively.


%
\section{Two-mode Jaynes-Cummings model}
\label{sec:two-mode-JC}
\begin{figure*}[ht]
    \centering
    \includegraphics[width=\textwidth]{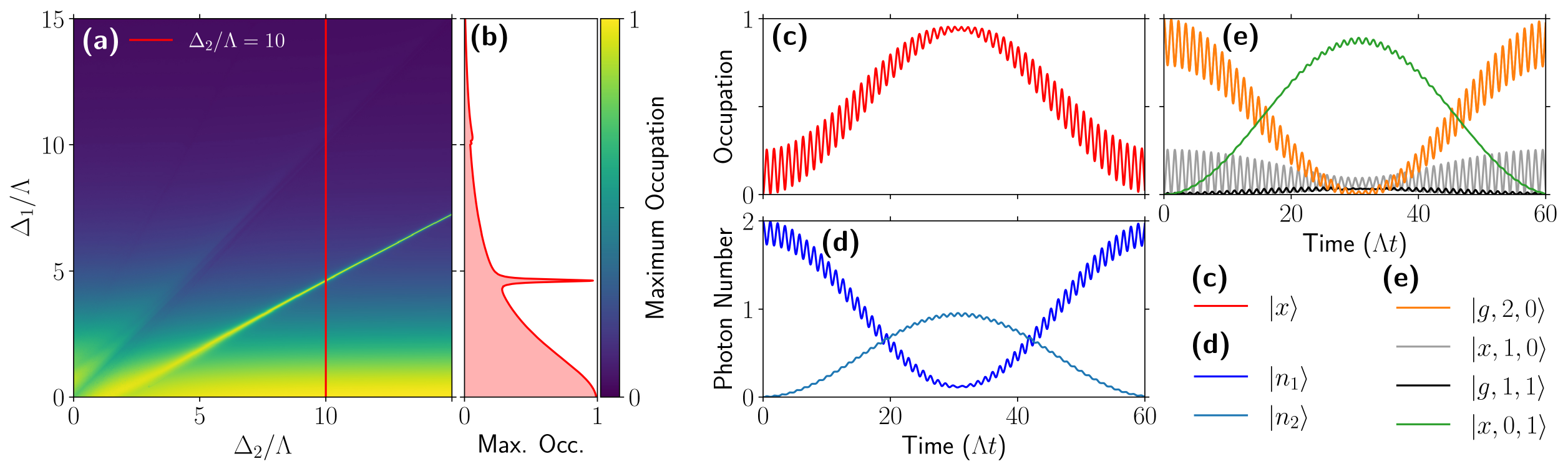}
    \caption{\textbf{Dynamics of a two-level system in the two-mode Jaynes Cummings model}: (a) Color map of the maximum occupation as a function of the detunings $\Delta_1$ and $\Delta_2$. The initial state is $|g, 2, 0\rangle$. (b) A cut for $\Delta_2=10\Lambda$ as indicated by the red line. Dynamics of the (c) excited state occupation and (d) occupation in the two photonic modes. (e) Dynamics of the occupations of selected states participating in the process. }
    \label{fig:two-photon-scan}
\end{figure*}

To describe the two-color excitation in a fully quantum mechanical picture, we describe the light field in terms of photon creation $(a_i^{\dagger})$ and annihilation $(a_i^{})$ operators, considering a two-mode Jaynes-Cummings Hamiltonian. In this system, the two-level system is simultaneously coupled to two photon modes $\omega_i$.  
We are focusing on the case of both modes being off-resonant with the two-level system 
\begin{equation}
    \omega_0 \neq \omega_1 \neq \omega_2 \neq \omega_0 \,.
\end{equation}
Such a configuration has been realized using atoms and two orthogonal cavities \cite{rauschenbeutel2001controlled}. There, the interaction between the two-level system and the two modes was controlled by step-wise changing of the detuning \cite{rauschenbeutel2001controlled,gonta2009generation}.

Another realization of the two-mode Jaynes-Cummings model could be via superconducting circuits \cite{mariantoni2008two, neeley2009emulation,rosario2023collateral}, where the frequency of the modes can be adjusted by changing the capacitance or inductance of the resonators.

The corresponding Hamiltonian reads
\begin{eqnarray}
    {H} &=& \hbar \omega_0 |x\rangle\langle x| 
        + \sum_{i=1}^2 \hbar \left[ \omega_i {a}^\dag_i{a}_i
     + \Lambda_i (|g\rangle\langle x|{a}^\dag_i + |x\rangle\langle g|{a}_i)  \right] \notag \\
    &=& \hbar \omega_0 \sigma_+\sigma_-
        + \sum_{i=1}^2\hbar \left[ \omega_i {a}^\dag_i{a}_i +  \Lambda_i (\sigma_- {a}^\dag_i + \sigma_+ {a}_i) \right]   \,.
\end{eqnarray}  
The coupling strength between the two-level emitter and mode $i$ is given by the parameter $\Lambda_i$. These coupling strengths translate into the Rabi frequencies, which depend on the total number of excitations in the coupled emitter-cavity system. While in the first row we use the notation commonly used in the quantum dot community, in the second row we have rewritten the equation using Pauli matrices, as commonly used in the atomic physics community.

We emphasize that our focus is different to the case of two resonant modes, i.e., $\omega_1=\omega_2=\omega_0$ or two off-resonant, but degenerate modes with $\omega_1=\omega_2\neq\omega_0$. These cases have been studied previously \cite{Larson2021book, rosario2023collateral,papadopoulos1988energetics, benivegna1994new, benivegna1996quantum,xie1995photon}. For quantum dots, the case of two resonant modes with different polarization has been realized \cite{meier2004spin}. We note that our model is distinct from the two-photon Jaynes-Cummings-model with bi-modal coupling, where the de-excitation leads to the direct emission of two photons ($\sim \sigma_- {a}^\dag_1 {a}^\dag_2$) \cite{guo1993characteristic,guo1989quantum}, which has been analyzed in the context of squeezing \cite{duan2015solution}.

We compose the basis $|\nu , n_1 , n_2\rangle$ as product states of the electronic states $|\nu\rangle \in \{|g\rangle, |x\rangle \}$ and the Fock states $|n_i\rangle$ in mode $i$. The interaction between the two-level system and the photon modes is either allowed by absorption of a photon coupling the states, meaning
\begin{subequations}
\begin{eqnarray}
    |g, n_1, n_2\rangle \leftrightarrow |x, n_1-1, n_2\rangle,\\
    |g, n_1, n_2\rangle \leftrightarrow |x, n_1, n_2-1\rangle,
\end{eqnarray}
or emission of a photon, if one reads the equations from right to left.
\end{subequations}
We emphasize that it does not allow for a direct photon exchange between the two modes, but all interactions are mediated via the two-level system. 

\section{Dynamics and resonance conditions}
\label{sec:dynamics}

To study the conditions under which few-photon scattering might lead to an excitation of the emitter, we analyze the dynamics of the two-level system coupled via a given number of initially present photons and search for the maximal occupation of the excited state. To this end, we set up the equations of motion with the initial state of the two-level system being in the ground state and a given number of photons as  $|g,n^{(0)}_1,n^{(0)}_2\rangle$. Defining the operator for the total number of excitations as $\hat{N} = \sum_j {a}^{\dag}_j {a}^{\phantom{\dag}}_j + |x\rangle\langle x|$, it is easily seen that $N^{\text{tot}}=\langle\hat{N}\rangle$ is conserved in the Jaynes-Cummings model. Our initial conditions thus restrict the dynamics to eigenstates of $\hat{N}$ with total excitation number $N^{\text{tot}}=n_1^{(0)} + n_2^{(0)}$, which is a manifold with finite dimension.

Diagonalization of the two-mode Jaynes-Cummings Hamiltonian for given initial conditions provides a simple solution of the time-dependent Schrödinger equation via the time-evolution operator. From the dynamics, we obtain the maximally achievable occupation of the excited state as displayed in the results. Additionally, we checked our results by direct numerical integration via the free software packet QuTip \cite{qutip}. 

For a single photon in either of the two modes, i.e., $|i\rangle = |g, 1, 0 \rangle$ or $|i\rangle = |g, 0, 1 \rangle$, according to the Rabi condition for a two-level system, the occupation of the excited state remains small for $\Delta/ \Lambda \gg 1$. Here, we are interested in large occupations for detunings in this parameter regime, where for the initial condition of a single photon no significant excited-state occupation is reached. 

\subsection{Two-photon scattering}
    \label{sec:twophotonscattering}
    
The minimal number of photons required to achieve a non-negligible excitation of the two-level emitter are two photons in the mode with the smaller absolute value of detuning. This corresponds to the initial state
\begin{equation}
    |i\rangle = |g, 2, 0 \rangle
\end{equation}
We scan the maximally achievable occupation of the excited state in the two-level system as a function of the two detunings $\Delta_1$ and $\Delta_2$ as shown in Fig.~\ref{fig:two-photon-scan}(a), while setting $\Lambda=\Lambda_1=\Lambda_2$. A cut for $\Delta_2=10\Lambda$ is shown in Fig.~\ref{fig:two-photon-scan}(b). 

If the emitter is coupled to only a single mode, the maximally achievable occupation of the excited state would show a Lorentzian-shaped decrease for increasing detuning $\Delta_1$. In the color map, this is seen when going from bottom to top where the color fades from yellow (one) to blue (zero).

Most strikingly, a sharp maximum of the excited-state occupation is found on top of the Lorentzian decrease. This sharp maximum reaches almost unity. In the color map, this is observed as a clear line with a slope of about $\Delta_1 \approx \tfrac{1}{2}\Delta_2$. This line matches the resonance condition where the SUPER mechanism enables the inversion of the two-level system. In the cut in Fig.~\ref{fig:two-photon-scan}(b) we also see the sharp resonance appearing at $\Delta_1/\Lambda=4.62$.

To understand what is happening at the resonance line, we look at the exemplary dynamics of different quantities shown in Fig.~\ref{fig:two-photon-scan}(c-e). Fig.~\ref{fig:two-photon-scan}(c) shows the dynamics of the occupation of the excited state $\big\langle |x\rangle \langle x| \big\rangle $ alongside the dynamics of the photon mode occupation $n_i = \langle a_i^{\dagger} a_i^{}\rangle$ in Fig.~\ref{fig:two-photon-scan}(d). The excited state occupation exhibits a fast oscillation on top of a slower oscillation and, eventually, the system reaches inversion. This behavior reminds of the semiclassical SUPER case [cf. Fig.~\ref{fig:semi}(b)]. Surprisingly, the excitation is achieved just by coupling the system to the second mode, though the latter holds no initial photon.

From the dynamics of the photon numbers it becomes clear that the two photons in mode $1$ both vanish, i.e., $|n_1=2\rangle \to |n_1=0\rangle$. Simultaneously with the excited-state occupation, the photon mode $2$ becomes occupied and holds a single photon with $|n_2=1\rangle$ at its maximum. We also see that the fast oscillation mostly takes place only in the photon mode $|n_1\rangle$. 

According to the coupling rules in the Jaynes-Cummings model, the transition from the initial state $|i\rangle=|g, 2, 0\rangle$ to the final state $|f\rangle=|x, 0, 1\rangle$ does not take place directly, but via
\begin{equation}
|i\rangle=|g, 2, 0\rangle \leftrightarrow |x, 1, 0\rangle \leftrightarrow |g, 1, 1\rangle \leftrightarrow |x, 0, 1\rangle =|f\rangle
\end{equation}
For the conditions required for achieving maximal excited-state occupation, the initial state $|i\rangle$ and final state $|f\rangle$ are almost degenerate, leading to a resonant transfer, whereas the other states are mostly unoccupied during the dynamics. This is confirmed in Fig.~\ref{fig:two-photon-scan}(e), where the dynamics of the occupation of these states is shown. Indeed, the occupations of the intermediate states remains small. On top of the overall dynamics, a fast oscillation is observed as a signature of the SUPER mechanism.

In the SUPER scheme in the semiclassical picture \cite{bracht2021swing, bracht2023dressed} the resonance condition (cf. Eq.~\ref{eq:resonance_2CSUPER}) was derived as 
\begin{equation}
    \Delta_2 = \Delta_1 + \sqrt{\Delta_1^2 + (\Omega_1^{\text{max}})^2} \,\,
    \stackrel{\Delta_1 \gg \Lambda_i}{\longrightarrow}  \, \, 2\Delta_1 \,.
\end{equation}
A similar resonance condition has been predicted for the two-photon scattering in resonance fluorescence with  $\Delta_2=2\Delta_1$ in the limit of large detunings \cite{masters2023simultaneous,dalibard1983correlation}\,. 

An adiabatic elimination can be performed for the equations of motions of the two-mode Jaynes-Cummings model in order to derive an approximate resonance condition for large detunings 
\begin{equation}
    \Delta_2  \stackrel{\Delta_2\gg \Lambda}{\approx} 2\Delta_1 +{ 4\Lambda^2 \over    \Delta_1  }  
\end{equation}
The details of the calculation are given in App.~\ref{app:reduction}. This resonance condition agrees well with our results found in Fig.~\ref{fig:two-photon-scan}(a). In addition, similar results are found in both the semiclassical and quantum pictures. This corroborates our assumption that the Jaynes-Cummings model with two off-resonantly coupled modes is indeed the adequate quantum mechanical analog of the semiclassical SUPER scheme.

It is interesting that these findings are connected to the idea of two-photon scattering which was recently observed in experiments \cite{masters2023simultaneous,le2022dynamical} in resonance fluorescence. 

We note that in the area close to the diagonal line, where both modes are degenerate with $\Delta_1=\Delta_2$, there is a bump observable in the cut as well. The degenerate case has been discussed in the literature \cite{papadopoulos1988energetics, benivegna1994new, xie1995photon, benivegna1996quantum}, as it shows different physics compared to the coupling to a single mode. The bump appears in the vicinity of the diagonal line including several small features, that we attribute multi-photon processes. Since here we are interested in the off-resonant case where both modes are distinct from each other and detuned to the two-level quantum emitter, we leave the detailed discussion of these processes to future research. 

\subsection{N-photon scattering}

\begin{figure}[htb]
    \centering
    \includegraphics[width=0.48\textwidth]{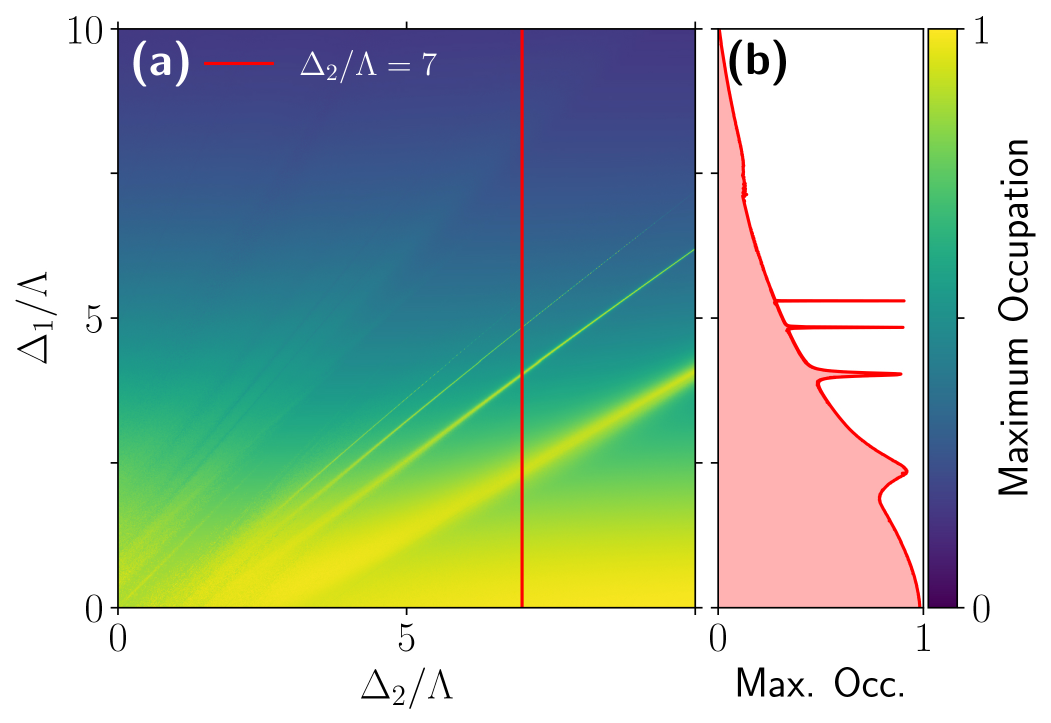}
    \caption{\textbf{Resonances for the multi-photon scattering}:(a) Color map of the maximal occupation of the excited state as a function of detunings $\Delta_1$ and $\Delta_2$ with (b) a cut at $\Delta_2/\Lambda=7$. The initial state is $|g, 5, 0\rangle$.}
    \label{fig:multi-photon-scan}
\end{figure}

Next, we consider the case of an initial state with 5 photons in one mode
\begin{equation}
    |i\rangle = |g, 5, 0 \rangle.
\end{equation}
The maximally achievable occupation of the excited state as a function of both detunings $\Delta_i$ is shown in Fig.~\ref{fig:multi-photon-scan}. On top of the single-mode Rabi background, we now find four distinct lines of maximal excitation probability. For example, for a cut at $\Delta_2/\Lambda=7$ we find resonances at $\Delta_1/\Lambda \approx 2.400,\, 4.039,\,4.839$ and $5.298$. Furthermore, we observe that the width of the resonance peaks decreases with increasing detuning.

Inspecting the lines, we find that they belong to multi-photon scattering processes effectively connecting the initial and final states
\begin{equation}
    |g, 5, 0\rangle \leftrightarrow |x, 5-N, N-1  \rangle 
\end{equation}
The line with the lowest slope belongs to the two-photon scattering processes $|g, 5, 0\rangle \leftrightarrow |x, 3, 1  \rangle$ discussed previously. The next line then features a three-photon scattering process with $|g, 5, 0\rangle \leftrightarrow |x, 2, 2  \rangle$. In this case, the five photons in the initial mode are distributed such that one photon excites the two-level system, two photons are scattered into the mode with higher detuning and two photons remain in the initial mode with lower detuning. In the curve with highest slope, we have a five-photon scattering process $|g, 5, 0\rangle \leftrightarrow |x, 0, 4  \rangle$. While one photon excites the two-level system, the other four photons are scattered into the higher detuned mode. 

The time required to achieve inversion increases for the higher-order processes as more photons $N$ are scattered from one mode to the other. This translates to a decreased effective Rabi frequency and is also reflected in the line widths that become narrower for higher $N$. We assume, that in an experiment that also includes dissipation, these processes are much less likely to occur and most likely cannot be stimulated.

\subsection{Scattering from both modes}
\begin{figure}[ht]
    \includegraphics[width=\columnwidth]{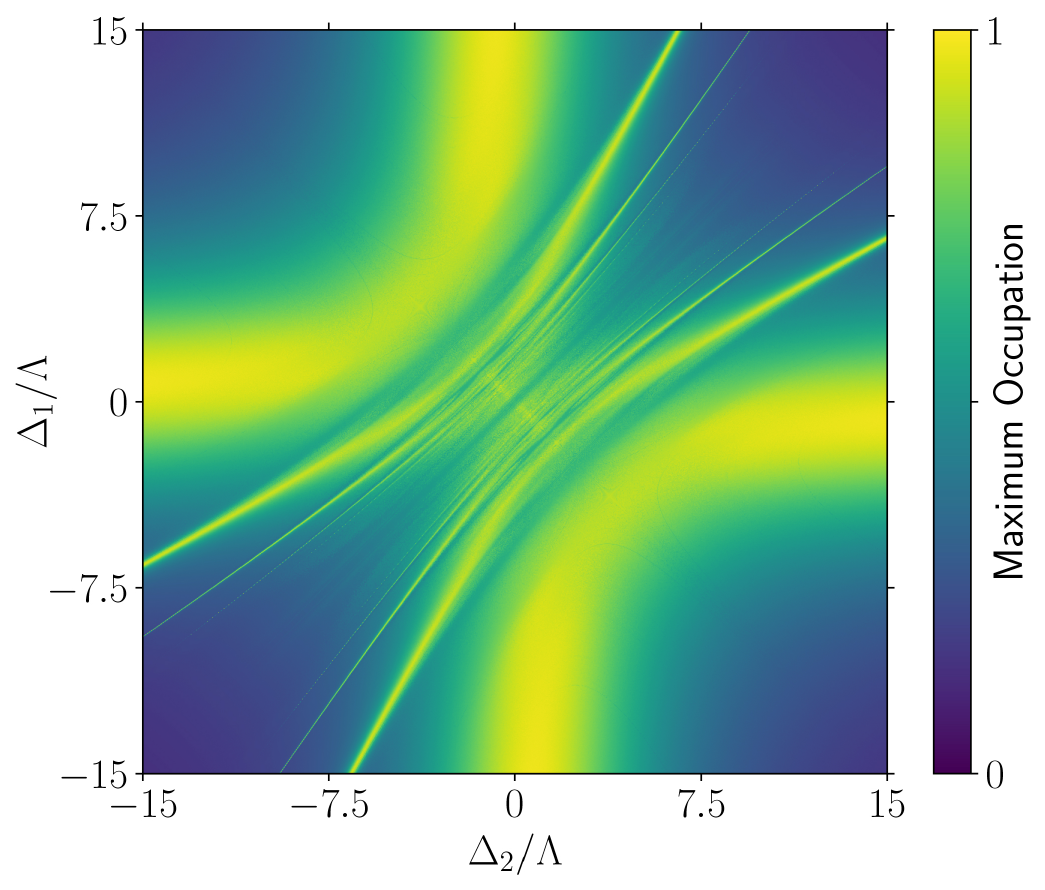}
    \caption{\textbf{Resonances for symmetric N-photon scattering}. Color map of the maximum occupation of the excited state as a function of detunings $\Delta_1$ and $\Delta_2$ for the initial state $|i\rangle = |g, 5, 5 \rangle$.}
    \label{fig:fullscan}
\end{figure}

In this section, we discuss the full scan of the maximally achievable occupation of the two-level emitter that is performed for the initial state 
\begin{equation}
    |i\rangle = |g, 5, 5 \rangle \,.
\end{equation}
In addition to both modes having negative detuning, we scan the region of positive detuning as well as the regions with one mode being positive and one mode being negative. The results are shown in Fig.~\ref{fig:fullscan}.

Several symmetries are apparent, as expected from our model: If we switch the sign of both detunings, the picture is symmetric. At the origin $\Delta_1=\Delta_2=0$, we do not achieve inversion as expected from the single-mode case. This is also found in Fig.s~\ref{fig:two-photon-scan} and \ref{fig:multi-photon-scan}. Here, the photon modes can be easily rewritten into two new modes that are coupled and uncoupled to the two-level system, respectively. Then, full inversion will only be reached if the coupled mode is initially in a pure Fock state with one or more photons. However, in our case the two coupled and uncoupled modes are initially entangled, resulting in a lower maximal occupation depending on the number of initially available photons. 

We also see distinct resonance conditions for both detunings being either negative or positive. Though we now also get resonances for $\Delta_2>\Delta_1$, in all cases photons are scattered from the lower detuned mode to the higher detuned mode. Note that in the case with non-equal coupling constants $\Lambda_1 \neq \Lambda_2$ we get a similar behavior as shown in App.~\ref{app:asy}. 

Let us now turn to the the case $\Delta_1>0$ and $\Delta_2<0$ ($\Delta_1<0$ and $\Delta_2>0$). This case has been studied in the atomic community \cite{guerin2001dynamical,conde2005experimental}. In the quantum dot community it was named di-chromatic excitation \cite{he2019coherently, koong2021coherent}, and studied for single photon generation. It can be understood as a special case of the SUPER scheme \cite{bracht2023dressed}.

For the detunings having opposite signs, we find a broad area where the  excited-state occupation reaches close-to-unity values. From the di-chromatic SUPER case \cite{bracht2023dressed}, the predicted resonance condition would be
\begin{equation}
    \Delta_2 = |\Delta_1| - \Omega^{\text{Rabi}} 
\end{equation}
with $\Delta_1$ belonging to the initially occupied mode and $\Omega^{\text{Rabi}}$ is the generalized Rabi frequency in the classical picture. Accordingly, the second detuning $\Delta_2$ is rather small in agreement with the broad area lying close to zero detuning. 

\section{Discussion and conclusions}
\label{sec:conclusion}

\begin{figure}
    \centering
    \includegraphics{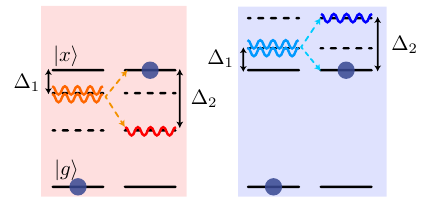}
    \caption{\textbf{Sketch of the photon-photon scattering:} Illustration of a two-level system that interacts with two photon modes with detunings $\Delta_{1,2}$, either red- or blue-detuned to the transition of the two-level system. Initially, one mode holds two photons, which results in an occupation of the two-level system and a photon in the other mode.}
    \label{fig:Auger}
\end{figure}

In summary, we have found a scattering mechanism in the two-mode Jaynes Cummings model, where the electronic system is promoted from its ground state to its excited state, while off-resonant photons are transferred from one mode to another according to
\begin{eqnarray}
    &&|g,n_1,n_2\rangle   \leftrightarrow |x,n_1-N,n_2+(N-1)\rangle .
\end{eqnarray}
We found that $N$, with $N\ge 2$, photons scatter in such a way that one photon excites the two-level emitter while the other $N-1$ photons scatter into the second off-resonant mode. This second off-resonant mode must at least have twice the detuning of the lower detuned mode.

Our findings relate to different results in quantum optics: They are in line with the observations of multi-photon scattering processes in resonance fluorescence, predicted theoretically \cite{dalibard1983correlation} and recently observed experimentally \cite{masters2023simultaneous}. Going beyond the two-photon scattering, we furthermore discussed that also scattering processes including three or more photons are in principle possible, but harder to realize in experiment. 

The dynamics and resonance conditions are also in line with the SUPER scheme, which was found for laser pulses described by classical light fields \cite{bracht2021swing,karli2022super}. In the SUPER scheme, full inversion became possible only when two laser pulses with different detunings were applied to the two-level system. This corresponds to stimulating the two-photon process described in the present manuscript. Note that in SUPER the excitation is usually performed with finite laser pulses, while in Fig.~\ref{fig:semi} we considered the case of an excitation with constant amplitude that is switched on instantaneously. For smooth pulses, adiabatic undressing effects \cite{Barth2016} may push the occupation to its maximum.

We have depicted the two-photon scattering process in Fig.~\ref{fig:Auger}. In the red-detuned case, the scattering leads to the creation of photons with lower energy, while the photon scatterings for blue-detuned photons leads to an up-rise in photon energy. The process also reminds of Auger-scattering, which in the solid-state community is known to occur only between electrons, but recently in quantum dots also the radiative Auger process, where two electrons scatter accompanied by the emission of a photon, was demonstrated \cite{lobl2020radiative,yan2023coherent}.  Our analysis provides a deeper understanding for the SUPER scheme at the fully quantized level. The results show that a SUPER-type inversion of a quantum emitter is not only possible at high laser intensities where the semiclassical description applies, but can in principle already be achieved with as few as two off-resonant photons provided a second off-resonant mode is available. \footnote{After submission of our manuscript, a similar work was published \cite{vannucci2024super}.}

The quantum emitter inversion enables a photon scattering between off-resonant modes that might lead to new applications in quantum photonics like frequency conversion or photon addition/subtraction. In order to describe these processes in a realistic system, one needs to account for dissipative effects. For making the effects observable the coupling strengths to both modes have to exceed the dissipation rates. For example, in a quantum dot system, the electron-phonon interaction is a prominent source of dissipation \cite{Reiter2019} and also affects the SUPER scheme \cite{bracht2022phonon} even for two red-detuned lasers. Dissipative effect will lead to an asymmetry between positive and negative detunings. Additionally, one should account for cavity losses. Therefore, we assume that in an experiment the photon scattering with $N\ge3$ will be extremely difficult to observe. In the case of superconducting circuits, also the coupling between the bosonic modes can become of importance, introducing new dephasing channels \cite{mariantoni2008two}. . For simulating experiments on such systems, dissipative effects should be included, which we leave for future work. We are confident that with the advancements in solid-state cavities \cite{schneider2016quantum,androvitsaneas2023direct} the presented effects should be observable in solid-state quantum emitters.

\section{Acknowledgements}
AR acknowledges funding by the Alexander von Humboldt Foundation in the framework of the Alexander von Humboldt Professorship endowed by the Federal Ministry of Education and Research (Germany).

%

%

%
%
\appendix

\begin{widetext}

\section{Reduction to a two-level system}\label{app:reduction}

\begin{figure}
    \includegraphics[width=0.48\textwidth]{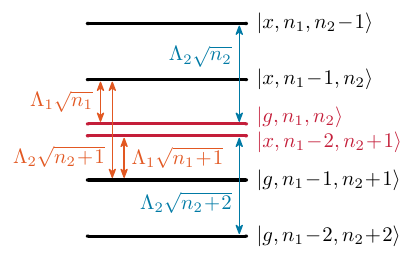}
    \caption{\textbf{Scheme of the most relevant states:} Display of the most relevant states for coupling between the initial state $ |i\rangle=|g,n_1,n_2\rangle$ and $|f\rangle =|x,n_1-2,n_2+1\rangle$.}
    \label{fig:SW}
\end{figure}

As discussed in the main text, at the discussed resonance conditions the dynamics effectively takes place usually between only two states, namely
\begin{equation}
    |i\rangle =|g, n_1, n_2\rangle \leftrightarrow |f\rangle = |x,  n_1-N, n_2+(N-1)  \rangle 
\end{equation}
For the condition that
\begin{equation}\label{eq:approx}
 \Lambda_i\sqrt{n_i}\ll |\Delta_1-\Delta_2|,\    
\end{equation}
the system can be reduced to the most relevant states $  |1\rangle = |x,n_1,n_2-1\rangle,   |2\rangle= |x,n_1-1,n_2\rangle,|3\rangle= |g,n_1,n_2\rangle,|4\rangle=|x,n_1-2,n_2+1\rangle, |5\rangle=|g,n_1-1,n_2+1\rangle,|6\rangle=|g,n_1-2,n_2-2\rangle$ as indicated in Fig.~\ref{fig:SW}. We go into a rotating frame and shift the energy to be zero at the initial state. In this case the Hamiltonian reduces to
\begin{equation*}  \label{K-matrix}
H =
\hbar \left(\begin{array}{cccccc }
\Delta_2 &0 & \Lambda_2\sqrt{n_2} & 0 & 0 & 0  \\
0 &\Delta_1&  \Lambda_1\sqrt{n_1} &  0&\Lambda_2\sqrt{n_2+1}   &  0  \\
\Lambda_2\sqrt{n_2}& \Lambda_1\sqrt{n_1}  &  0 &  0 & 0 & 0    \\
0& 0 &0   & 2\Delta_1-\Delta_2  &   \Lambda_1\sqrt{n_1-1} & \Lambda_2\sqrt{n_2+2}   \\
0& \Lambda_2\sqrt{n_2+1}  & 0 &  \Lambda_1\sqrt{n_1-1}  &\Delta_1-\Delta_2   &  0 \\
0&  0&  0& \Lambda_2\sqrt{n_2+2}  & 0&2\Delta_1-2\Delta_2    \\
\end{array}\right)
\end{equation*}
 With these states, the equations of motion for the coefficients of the state $|\psi\rangle = \sum_{i=1}^{6} C_i |i\rangle$ can be set up. From this, an adiabatic elimination can be performed to highlight the coupling between the states $|3\rangle = |i\rangle$ and $|4\rangle = |f\rangle$. In adiabatic elimination, the time derivatives of the coefficients $C_1,C_2,C_5,C_6$ are set to zero. The resulting equations can be further approximated on the same level as Eq.~\eqref{eq:approx} and yields expressions for $C_1,C_2,C_5,C_6$ as functions of $C_3,C_4$, such that closed equations of motion can be derived for the latter. This results in 
\begin{subequations}
\begin{eqnarray}
  i{d \over dt} C_3(t)&=& \left(  - {\Lambda_2^2 n_2 \over \Delta_2} - {  \Lambda_1^2  n_1  \over    \Delta_1 }\right)C_3(t)+ {  \Lambda_1^2\Lambda_2\sqrt{n_1-1} \sqrt{n_1} \sqrt{n_2+1}\over    \Delta_1(\Delta_1-\Delta_2)}C_4(t) ,\\
 i{d \over dt} C_4(t)&= & { \Lambda_1 ^2 \Lambda_2\sqrt{n_1-1}\sqrt{n_1} \sqrt{n_2+1} \over    \Delta_1(\Delta_1-\Delta_2 )} C_3(t)+\left[ 2\Delta_1 -\Delta_2 
-{ \Lambda_1^2( n_1- 1)\over    (\Delta_1-\Delta_2 )}  -{\Lambda_2^2(n_2+2) \over 2( \Delta_1 -\Delta_2)}\right]C_4(t)).
\end{eqnarray} \end{subequations}
Now, we have reduced the system to an effective two-level system described by the effective Hamiltonian
\begin{equation}
 H^{\text{eff}} = \hbar \left(\begin{array}{cc}
E_1 &\frac{\Omega^{\text{eff}}_R}{2}\\
\frac{\Omega^{\text{eff}}_R}{2}& E_2\\
\end{array}\right),
\end{equation}
introducing the effective Rabi frequency
\begin{equation}
    \Omega_R^{\text{eff}} ={  2\Lambda_1^2\Lambda_2\sqrt{n_1-1} \sqrt{n_1} \sqrt{n_2+1}\over    \Delta_1(\Delta_1-\Delta_2)} 
\end{equation}
and the effective energies of the two states including Stark-Shifts
\begin{subequations} \begin{eqnarray}
E_1 &=&
    - {\Lambda_2^2 n_2 \over \Delta_2} - {  \Lambda_1^2  n_1  \over    \Delta_1 },\\
E_2 &=& 2\Delta_1- \Delta_2 +
    { \Lambda_1^2( n_1- 1)\over    (\Delta_2-\Delta_1 )}  +{\Lambda_2^2(n_2+2) \over 2( \Delta_2 -\Delta_1)} .
\end{eqnarray} \end{subequations}
The resonance condition is given for $E_2-E_1=0$, which assuming $2\Delta_1-\Delta_2\approx 0$, is approximated by

\begin{equation}
    \Delta_2 =2\Delta_1 +{ \Lambda_1^2\over    \Delta_1  }( 2n_1- 1)  +{\Lambda_2^2 \over \Delta_1 }  (n_2+1).
\end{equation}
In the case discussed in Sec.~\ref{sec:twophotonscattering} we have $n_1=2$ and $n_2=0$ as well as $\Lambda_1=\Lambda_2=\Lambda$. Under these circumstances, the resonance condition reads
\begin{equation}
    \Delta_2 =2\Delta_1 +{ 4\Lambda^2 \over    \Delta_1  }  .
\end{equation}
We remind that this describes the behavior for detuning differences that are large compared to the coupling strength as given in Eq.~\eqref{eq:approx}. 

The same procedure of adiabatic elimination can be also performed for cases with $N\ge 3$. However, we these cases are much less likely to be observed, we present only the numerical results obtained in the main manuscript.

\section{Asymmetric coupling constants}\label{app:asy}

In the main manuscript we have always set the coupling constants of the two modes as equal. This might lead to the question, if this is a special situation. Thus, we analyze the case of unequal coupling constants in Fig.~\ref{fig:fullscan}(b), where we set the coupling constants to be $\Lambda_1 = \tfrac{1}{2}\Lambda_2$. The initial state is set to $|g, 5, 5 \rangle $ . As seen from the scan this introduces an asymmetry with respect to the anti-diagonal, e.g.,  the two-photon scattering $|g, 5, 5 \rangle \leftrightarrow  |g, 3, 1 \rangle$ is more pronounced than $|g, 5, 5 \rangle \leftrightarrow  |g, 1, 3 \rangle$. Nonetheless, the overall behavior is very similar, underlining that this is not an exceptional behavior just found for equal coupling constants.

\begin{figure}[htb]
    \includegraphics[width=0.5\columnwidth]{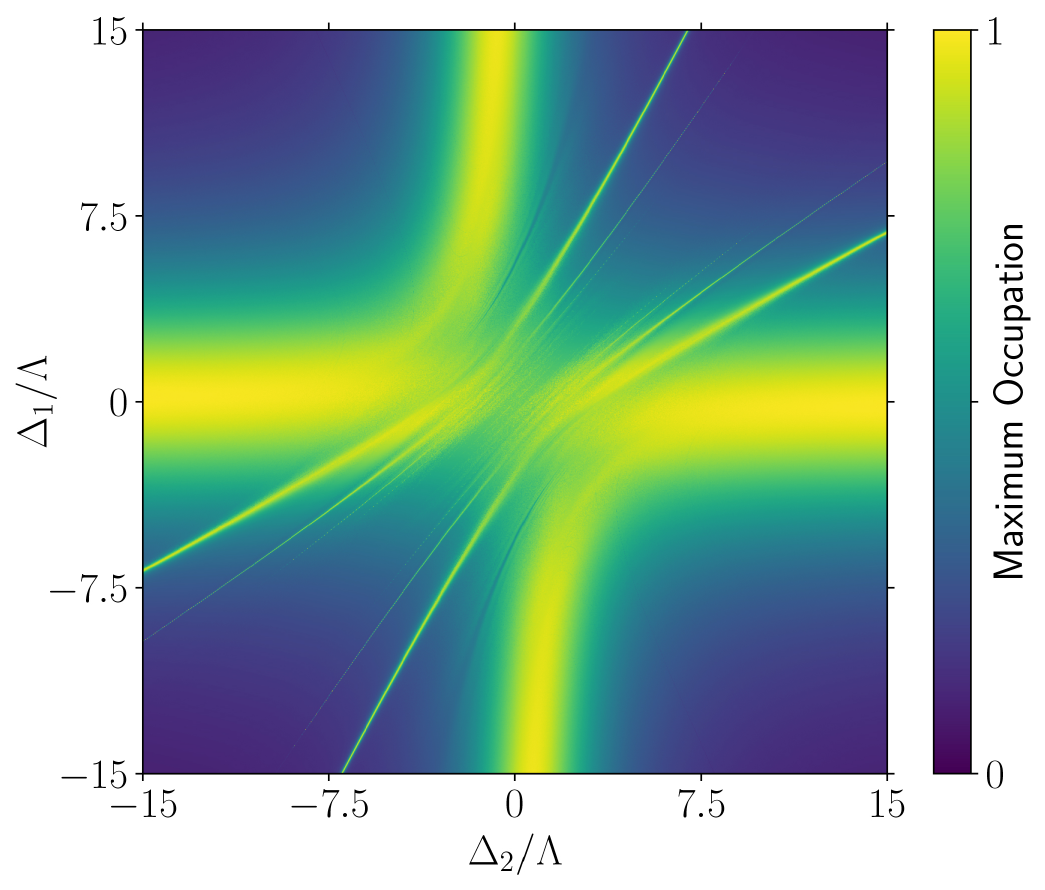}
    \caption{\textbf{Resonances for symmetric N-photon scattering}. Color map of the maximum occupation as a function of detunings $\Delta_1$ and $\Delta_2$ for the initial state $|i\rangle = |g, 5, 5 \rangle$ and unsymmetric coupling constants. We have set $\Lambda=\Lambda_1$ and $\Lambda_2=\tfrac{1}{2}\Lambda$.}
    \label{fig:fullscan_asy}
\end{figure}
\end{widetext}

\end{document}